\documentstyle[aps,eqsecnum]{revtex}

\newcommand{\be}{\begin{equation}}
\newcommand{\ee}{\end{equation}}
\newcommand{\bea}{\begin{eqnarray}}
\newcommand{\eea}{\end{eqnarray}}

\begin{document}

\draft
\title{Generalized Hamiltonian Dynamics of Friedmann
 Cosmology with Scalar and Spinor Matter Source Fields}

\author{
A. M. Khvedelidze\ $^a$ $^b$,\,\,
 and  \,
Yu.G.~Palii\ $^c$
}
\address{$^a$
A. Razmadze Mathematical Institute, Tbilisi, 380093, Georgia}
\address{$^b$
Bogoliubov Laboratory of Theoretical Physics,
Joint Institute for Nuclear Research, 141980 Dubna, Russia}
\address{$^c$
Laboratory of Information Technologies,
Joint Institute for Nuclear
Research, 141980 Dubna, Russia}

\date{\today}

\maketitle
\begin{abstract}
The classical and quantum dynamics of
the Friedmann-Robertson-Walker  Universe with
massless scalar and massive fermion  matter field as a source is discussed
in the framework of the Dirac generalized Hamiltonian formalism.
The Hamiltonian reduction of this constrained system is realized
for two cases of minimal and
conformal coupling between gravity and matter.
It is shown that in both cases for all values of curvature,
\(k= 0, \pm 1, \) of maximally symmetric space
there exists a time independent reduced local Hamiltonian
which describes the dynamics of the cosmic scale factor.
The relevance of conformal time-like Killing vector fields
in FRW space-time to the existence of time independent Hamiltonian
and the corresponding notion of conserved energy is discussed.
The extended quantization with the Wheeler-deWitt equation
is compared with the canonical quantization of unconstrained system.
It is shown that quantum observables treated as expectation values of
the Dirac observables properly describe the original classical theory.
\end{abstract}



\section{Introduction}


Cosmological models apart from the main task, to investigate
the large scale structure of the Universe, are highly
attractive  objects with the standpoint of analysis of the
conceptual problems in the theory of gravitation.
By studying  cosmological models  instead of general spacetime
we can to overcome the difficulties due to  the
infinite number of degrees of freedom and  concentrate attention
to the problems arising solely from the
time reparametrization invariance; such as the construction of
observables.
\footnote{The problem of observables  consist in the determination
of the invariant characteristics of gravitational field in
terms of measurable quantities \cite{Bergmann} and
is closely related to that of time evolution
\cite{KucharTime}, \cite{HajicekTime}.}
In the present article we attempt a contribution to
the discussion of some aspects  of this problem by
considering the simplest cosmological model,  the
Friedmann-Robertson-Walker (FRW) Universe filled
in the scalar
massless and massive spinor matter fields.
The conventional  Hamiltonian description  of this model
is based on the original Dirac \cite{Dir58} and the
so-called Arnowitt-Deser-Misner (ADM) \cite{ADM} formulation
of general relativity.
\footnote{
For review of the cosmological
models  construction  with  applications of the ADM method see e.g.
\cite{Ryan}.}
The ADM method involves the choice of certain coordinate
fixing conditions (gauge),
solution of the constraints and construction of the observables
such as energy, momentum and  angular momentum, using the
asymptotically flat boundary condition for gravitational field
and assuming that three-dimensional space of constant time is open
\cite{ReggeTeit}.
However,  when the closed Universe is considered
to build the ADM observables from initial data
for canonical variables it is impossible. Since in this case
there is no boundary of the space manifold and no
asymptotic region can  be used to construct the corresponding
integrals of motion. This leads to the conclusion that
for such cosmological models neither the natural notion of time
evolution nor the corresponding energy definition is possible
to find \cite{Faddeev}, \cite{Torre}.
To clear up this contradiction between the existence of widely used
cosmological quantities  and their  absence in the  corresponding
field theoretical formulation the FRW cosmological model will be
considered in the framework of the Dirac Generalized Hamiltonian
formulation \cite{DiracL}-\cite{HenTeit}.
The key moment of the canonical treatment is
the assumption that  general relativity represents ``already
parametrized'' theory due to the principle of general
covariance, so that the problem of construction
of observables  can be solved automatically  rewriting the
theory in the equivalent ``deparametrized'' form.
\footnote{To make  agreement between the four-dimensional covariance
and the possibility to extract from the canonical
coordinates hidden  variables appropriate for deparametrization theory
is difficult task.   To solve this problem
Kuchar suggested to perform  the ``second parametrization'' of general
relativity by extending its  phase space by the additional embedding
variables  \cite{KucharEmbed}.}
However, careful analysis of correctness of such
deparametrized program carried out by  Hajicek \cite{HajicekTop}
shows that even for
simple mechanical system with one quadratic Hamiltonian constraint
there are topological obstructions to its implementation
analogous of the well-known ``Gribov ambiguity''
in gauge theories.
A direct way to clarify the topological structure of such a
theory lies in the finding of integral curves
of the dynamical equations and
the investigation their  global properties.
Within this motivation the present note is devoted to the
realization of local deparametrization of integrable cosmological
FRW models considering it as a preparation for the study the global
features of reduction procedure.
\footnote{Apart from topological obstruction arising due to the
projection onto the  constraint  shell it is necessary
also to investigate the
problems connected with the topological structure  of
spaces of constant curvature. The well elaborated classification of
 three-dimensional spacelike manifolds \cite{Wolf} allows to
 estimate the influence of topological properties on
physical quantities.
An interesting study of the role played by this global properties
is under present consideration (see \cite{Bernui} and references therein).
}
We will follow the method  of Hamiltonian reduction
to construct the  observables and
the corresponding dynamical equations which is well elaborated
for  gauge theories. This approach is  based on an appropriate
choice of canonical coordinates on phase space and deals without
explicit introduction of any gauge-fixing functions
(see e.g. \cite{GKP} and references therein).

The general plan of present article is as follows.
In Section II we state the FRW cosmological model
with real massless scalar and massive fermion fields as sources with
different type of coupling to gravity.
In Section III some generic features of the Hamiltonian
reduction and the construction of observables in reparametrization
invariant mechanical models is disscused.
The aim of Section III
is to explain the method to  obtain  the unconstrained system from
reparametrized invariant one by considering the simplest example
of free  relativistic particle motion.
Section IV is devoted to the construction of
the unconstrained systems equivalent
to FRW cosmology when the homogeneous matter is presented in
different forms: as
massless scalar field interacting  with gravity minimally and
conformaly, massive spinor field.
Finally, in Section V we discuss the correspondence principle
fulfillment for observables in quantum theories based either on
Wheeler-deWitt equation or on the canonical quantization scheme
of  the unconstrained classical system.
In Appendices  we state some notations and technical details
of derivations in order to simplify the reading of the main text.


\section{Model with spatial homogeneity and isotropy}


By definition, the FRW spacetime  is a four-dimensional
pseudo-Riemannian manifold on which a six-dimensional Lie group \(G_6\)
acts as group of isometries. The group of isometries \(G_6\) has a
three-dimensional isotropy subgroup
and three-dimensional subgroup which acts simply transitive on
the one parameter  (``time $t$'')  family of spacelike hypersurfaces
\( \Sigma_t \).  The large group of isometries restricts both the
dependence and the number of independent components of the metric tensor
and leads to the so-called  maximally symmetric three-dimensional space.
After the choice of standard coordinates \cite{MTW}
one has the FRW metric
\begin{equation}  \label{eq:FRWmetric}
ds^2 = - N^2(t)\, dt \otimes dt +
a^2(t)\, \gamma_{ab} \, d{x}^a \otimes d{x}^b ,
\end{equation}
where  \( \gamma_{ab}\) is the time independent metric of
three-dimensional space
\begin{equation}
\gamma_{ab}\, d{x}^a \otimes d{x}^b = \frac{dr^2}{1-\frac{kr^2}{r_o^2}} +
  r^2(d\theta^2 + \sin^2\theta d\varphi^2)
\end{equation}
of constant curvature
\(
{}^{(3)}R(\gamma_{ij})=-6k/r_o^2, \quad k = 0, \pm 1  \).
The  lapse  function \( N(t) \)  and the cosmic scale factor \( a(t) \)
describe the remaining gravitational  degrees of freedom whose
classical behavior is determined by varying the
standard Hilbert action. However, constructed in this way
the minisuperspace model is out of interest. Simple counting of the
physical degrees of freedom shows that this  vacuum
FRW model is empty on the classical level; only unphysical degrees of
freedom propagate. Thus in order to have some  nontrivial observables it is
necessary to introduce the source matter fields.

\subsubsection{Lagrangian for scalar field with minimal coupling to gravity
}

The introduction of a massless  scalar field as a source of gravity
results in  the simplest cosmological model which has direct
correspondence to the classical Friedmann model.
\footnote{Below we will point out the correspondence
the conventional Friedmann cosmology based on the Einstein equations
supplemented by certain matter equation of state.}
For a massless scalar field, the two most interesting couplings to gravity
extensively considered are the so-called minimal coupling and the
conformal one.
\footnote{It is well known that essentially all
types of couplings of free scalar field to the scalar curvature and its
kinetic term  can be reduced to minimal
coupling form using rescaling of the
metric and  scalar field redefinition \cite{Deser}.
In 1974 based on this type of transformations  Bekenstein \cite{Beken}
proposed the method of construction solution  for  particular
case of conformally coupled Einstein-scalar equation
from solution of the minimally coupled ones
(see also  \cite{CalColJack}).
The detailed investigation of this type solutions for FRW geometry
with spatial homogeneous  scalar  fields can be found in
\cite{Page}.
Note also the interesting consideration of the evolution of Friedman cosmology
driven by scalar fields, given in \cite{StarkovichCooperstock}
}

The Hilbert action for gravity minimally coupled to
massless scalar field
\begin{equation}    
W=\int d^4x \sqrt{-g}\left[-\frac{{}^{4}R}{2\kappa}+
      \frac{1}{2}g^{\mu \nu}\partial_\mu\Phi\partial_\nu\Phi\right]
\end{equation}
reduces to the following
\begin{equation} \label{tru}
W=V_{(3)}
\int dt \left[-\frac{3}{\kappa} \left(
\frac{\dot{a}^2}{N_c} - \frac{ka^2}{r^2_o} N_c \right) +
\frac{a^2}{2N_c} \dot{\Phi}^2
 + \frac{3}{\kappa}
\frac{d}{dt}\left(\frac{a\dot a}{N_c}\right)\right]\;,
\end{equation}
assuming the spatial homogeneity  of the scalar field and
FRW metric (\ref{eq:FRWmetric}).
Here  \(\kappa=8\pi G\) and new variable $
N_c = {N}/{a}$ has been introduced. Integration over the
spatial hyperplane leads to the appearance of the factor
$V_{(3)}$ -- ``volume'' of the
three-dimensional space of constant curvature.
\footnote{In all formulas
this factor will be omitted, in order to simplify the
numerical factors.}

\subsubsection{Lagrangian for a scalar field with
conformal coupling to gravity }

The conformally coupled scalar field is described by action
\begin{equation} \label{eq:lagrG+M}
W[ g, \Phi ] = \int\limits_{}^{}d^4x\sqrt{-g}\left[-\frac{1}{2\kappa}\,
{}^{(4)}R
  + \frac{1}{12}\, {}^{(4)}R\Phi^2 + \frac{1}{2}
g^{\mu\nu}\partial_\mu\Phi\partial_\nu\Phi\right].
\end{equation}
Choosing the  metric (\ref{eq:FRWmetric}) this leads
to the action  for the FRW Universe filled in by massless
homogeneous scalar field \({\varphi(t)}:= {a(t)}\Phi (t)\)
\begin{eqnarray}  \label{eq:lagrFWW+MS}
&&W[a, N_c, \varphi] = \int dt\left[-
\frac{3}{\kappa}\left(\frac{\dot a^2}{N_c}-
\frac{ka^2}{r_o^2}N_c\right)+ \frac{1}{2}\left(\frac{\dot \varphi^2}{N_c}-
\frac{k\varphi^2}{r_o^2}N_c\right) +
\frac{3}{\kappa}\frac{d}{dt}\left(\frac{\dot a a}{N_c}\right)\right]\,.
\end{eqnarray}

\subsubsection{FRW Lagrangian with spinor matter fields}

The combined system of Dirac field and FRW metric have been investigated
from classical and quantum point of view by many authors.
In present article we explore the model closely related to the
formulation given in ~\cite{Isham,Zanelli}.
The starting point is
the action for a massive spinor field interacting with
gravity is given by
\begin{equation}      \label{EH}
W=\int d^4x\sqrt{-g}\left[-\frac{{}^{(4)}R(g)}{2\kappa}+
\frac{i}{2}\left(\bar\Psi\gamma^\mu(x)\bigtriangledown_\mu\Psi-
\bigtriangledown_\mu\bar\Psi\gamma^\mu(x)\Psi\right)
-m\bar\Psi\Psi\right],
\end{equation}
where the spinor  field $\Psi(x)$
( $\bar\Psi$ Dirac conjugate spinor field) components are treated
classically as a collection of
Grassmann variables  \(\Psi_i\Psi_j +\Psi_j\bar\Psi_i=0 \) and
$\bigtriangledown_\mu$ is the covariant derivative (see notation in Appendix \ref{appA}.).
Assuming the homogeneity of the  fermion fields and after the
redefinition $\psi(t):=a^{3/2}(t)\Psi(t) $
Eq.(\ref{EH}) reduces to the action of the finite dimensional system
\begin{equation}  \label{S}
W=\int dt\left[
-\frac{3}{\kappa}\left(\frac{\dot a^2}{N_c}-
\frac{ka^2}{r_o^2}N_c\right)
+\frac{i}{2}(\bar \psi\gamma^o\dot\psi
-\dot{\bar\psi}\gamma^o\psi)
- a N_c{\cal H}_D
+\frac{3}{\kappa}\frac{d}{dt}\left(\frac{a\dot a}{N_c}\right)
\right]\;,
\end{equation}
with
\begin{equation}  \label{HD}
{\cal H}_D=m\bar{\psi}{\psi}.
\end{equation}


\section{Reduction and observables in reparametrization invariant
mechanical models}


It is the purpose of this part to discuss  the  construction of
observables for a system with reparametrization invariance.
For our aims we shall state the ideas
using a mechanical system, i.e.  a system  with a finite number of degrees
of freedom and restrict ourselves to the case of Abelian constraints.

Let us consider a system with \( 2n \) - dimensional
phase space \(\Gamma \)  whose dynamics is constrained to
a certain submanifold  \(\Gamma_c \) describing by the
functionally independent set of \( m \) abelian constraints
\begin{equation}  \label{eq:constr}
\varphi_\alpha (p,q) = 0, \quad \quad  \quad \quad
\{ \varphi_\alpha (p,q), \varphi_\beta  (p,q)\} = 0\,.
\end{equation}
Due to the presence of constraints the Hamiltonian dynamics is
described by the Poincar\'e-Cartan form
\begin{equation}  \label{eq:P-C}
\Theta =  \sum_{i=1}^{n} \, p_i dq_i -  H_E (p,q) dt  \,,
\end{equation}
with the extended
Hamiltonian \(H_E (p,q) \) differing from the  canonical Hamiltonian
\( H_C (p,q)\)  by  a linear combination of  constraints with arbitrary
multipliers \( u_\alpha (t) \)
\begin{equation}
H_E (p,q) = H_C (p,q) + u_\alpha (t) \varphi_\alpha (p, q) \,.
\end{equation}
For the case of first class constraints the functions
\( u_\alpha (t) \) can't be fixed without using some additional
requirements.
This observation reflects the  existence of the local (gauge) symmetry
and the presence of coordinates in the theory whose  dynamics
is governed in an arbitrary way.
However, according to the principle of gauge invariance, these coordinates
do not affect  physical quantities and thus
can be treated as ignorable (gauge degrees of freedom).
The question is how to  identify these coordinates.
If theory contains only  Abelian constraints
one can find these ignorable coordinates as follows.
It is always possible \cite{Levi-Civita} -
\cite{Shanmugadhasan} to  define a canonical transformation to a new
set of canonical coordinates
\begin{eqnarray} \label{eq:cantr}
q_i & \mapsto & Q_i = Q_i \left ( q , p \right ),\nonumber\\
p_i & \mapsto & P_i = P_i \left ( q , p \right ),
\end{eqnarray}
so  that  \( m \) of the new  momenta
(\(\overline{P}_1, \dots ,\overline{P}_m \))
become equal to the Abelian constraints
\begin{equation}
\overline{P}_\alpha = \varphi _\alpha (q, p)\,.
\end{equation}
In the new coordinates (\(\overline{Q}, \overline{P}\)) and
(\( {Q}^{\ast}, {P}^{\ast}\)) we have the following canonical equations
\begin{eqnarray} \label{eq:motion}
 \dot{Q}^{\ast} = \{{Q}^{\ast}, H_{Ph} \},
 &\quad \quad  \quad &  \dot{\overline{P}} = 0, \nonumber\\
 \dot{P}^{\ast} = \{{P}^{\ast}, H_{Ph} \},
 & \quad \quad  \quad & \dot{\overline{Q}} = u(t) ,
\end{eqnarray}
with the physical  Hamiltonian
\begin{equation} \label{eq:PhysHam}
 H_{Ph}(P^{\ast}, Q^{\ast}) := H_C(P, Q)\,
 \Bigl\vert_{ \overline{P}_\alpha = 0}.
\end{equation}
The physical Hamiltonian \(H_{Ph}\) depends  only on the \(( n-m )\)
pairs of new gauge-invariant canonical coordinates
\(( {Q}^{\ast}, {P}^{\ast} )\).
Moreover the  form of the canonical system  (\ref{eq:motion})
expresses the  explicit separation of the phase space into
physical and unphysical sectors.
Arbitrary functions \( u(t) \) enter only into the part
the equation for the ignorable coordinates \( \overline{Q}_\alpha \),
conjugated to the momenta \({\overline{P}}_\alpha\).
\footnote{This paper deals with Abelian constraints only,  but a few
remarks
on the general non-Abelian case may be in order.
A straightforward generalization to this situation is
unattainable; identification of momenta with constraints is
forbidden due to the non-Abelian character of constraints.
However, one can  replace the non-Abelian constraints by an equivalent
set of constraints  forming  an Abelian algebra and after this
implement the above mentioned Levi-Civita transformation.
For proofs of this Abelianization statement see e.g.
\cite{Sunder} -- \cite{HenTeit} and  the description of itterative
Abelianization conversion in \cite{GKP}.}

Trying to apply this program to any model with reparametrization invariance
we as a rule  reveal that the physical Hamiltonian defined by
(\ref{eq:PhysHam}) is zero and thus we have  the dynamics of
unconstrained system in the Maupertuis form
\begin{equation}  \label{eq:P-C-M}
\Theta_{Ph} = \sum_{i=1}^{n-m} \, P^\ast_i d Q^\ast_i  - d V ,
\end{equation}
where $dV$ is a total differential.
The problem is now how to deal with the zero Hamiltonian.
This situation in some sence opposite to the
case known from the Hamilton-Jacobi method of
integration of equations of motion.
The main idea of this method is to implement on the system with
Hamiltonian \( H(t,p,q) \) the canonical transformation
with generating function \( S(t, q,p) \), which is the solution for
the  equation
\begin{equation}
\frac{\partial S}{\partial t} + H\left(
 t, q, \frac{\partial S}{\partial q}\right) = 0\,.
\end{equation}
As a result the new Hamiltonian is zero and the
equation of motion in the new coordinates have the simplest form
\begin{equation}
 \dot{Q} = 0, \quad \quad \dot{P} = 0 \,.
\end{equation}
After reduction we have just a system in these coordinates
and the problem is to  reconstruct the nonzero Hamiltonian
in any other coordinates for the obtained uncontsrained system.
Two remarks to the picture described above
may be in order. There is no difference
between the local behavior in  systems obtained via the reduction
of reparametrization invariant theories.  The specific
properties, which make a difference of  systems are hidden in the total
differential in  the Poincar\'e-Cartan form.

Before passing to the construction  of  the reduced
phase space for FRW Universe it seems worth to
set forth  our approach to the same  problem of a free
relativistic particle.

\subsection{ Digress: Reduced dynamics of free
relativistic particle}

For the presentation of our procedure to construct the reduced dynamical
system from the degenerate system with reparametrization invariance
let us start with the simplest case of free motion of a particle in
Minkowski space-time  writing its  action in the form  close to
the cosmological Friedmann models (\ref{tru}),
(\ref{eq:lagrFWW+MS}), (\ref{S})
\begin{equation} \label{parac}
W[x,e] := \frac{1}{2} \int\limits_{T_1}^{T_2} d\tau
\left ( \frac{{\dot{x}}_\mu^2}{e} + e m^2 \right ).
\end{equation}
The independent configuration  variables are particle wordline coordinates
\(x_\mu (\tau)\) and the additional ``vielbein''
determinant \(e(\tau)\).

Invariance of the action (\ref{parac}) under the reparametrization of
time
\( \tau \rightarrow \tau'=f(\tau) \) spoils the uniqueness of
the Cauchy problem for the corresponding equations of motion.
Therefore the problem is to fix the part of the variables whose dynamics
will be unique and whose initial conditions  are free from any
constraints.
The usual way to deal with this problem consists in choosing of a gauge
which tights the parameter of evolution with the configuration variables.
For example, the proper time gauge fixing
\( x_0(\tau)  = \tau \) leads to the instant form of the dynamics for a
relativistic particle.
However, let us  act in  spirit of the previous section and try to
reproduce the results of the instant form of particle dynamics without
introduction of gauge conditions.

According to the Dirac prescription the generalized Hamiltonian dynamics
for the system (\ref{parac}) takes place on the phase space spanned by five
canonical pairs ($ e, p_e$) and ($ x_\mu ,p_\mu $)  restricted by
the primary constraint
\(
p_{e}=0\;
\)
and the secondary constraint
\begin{equation}
p_\mu p^\mu - m^2 = 0 \,.
\end{equation}
To take into account these constraints and to derive equations of motion
one can  consider the Poincare-Cartan 1-form
\begin{equation} \label{P-Cp}
\Theta: = p_e de+ p_\mu dx^\mu - H_T d\tau\;,
\end{equation}
with  the total Hamiltonian
\begin{equation}
H_T: =\frac{1}{2}e(p_x^2-m^2) + \lambda(\tau) p_e\;.
\end{equation}
The equation of motion together with both  constraints
follow from functional
\begin{equation}
W[e,p_e; x,p,; \lambda]: = \int \Theta\;,
\end{equation}
using independent variation of the canonical pairs $(e,p_e),(x,p)$
and the Lagrange multiplier $\lambda$
\begin{eqnarray}
&&\dot x_\mu = ep_\mu\;, \quad   \quad  \dot p_\mu=0\;, \\
&&\dot e=\lambda\;, \\ \label{eq:encon}
&& p^2-m^2 =0\;, \quad   \quad  p_e = 0.
\end{eqnarray}
Let us now convince ourselves that performing certain canonical
transformations
one can put the  equation in such form that the Lagrange multiplier
function $\lambda(\tau)$ enters only in the equation for one
canonical pair.  According to the general scenario
described in previous section each canonical transformation
$$
\left(\begin{array}{cc}
e     & p_e\\
x^\mu & p_\mu
\end{array}\right)
\longmapsto
\left(\begin{array} {cc}
e & p_e\\
X^\mu & \Pi_\mu
\end{array}\right)\;,
$$
that identifies one canonical momentum with the energy constraint
(\ref{eq:encon}),  say \(\Pi_0 \)
\begin{equation}
\Pi_0= \frac{1}{2}(p_x^2-m^2)
\end{equation}
leads to this pattern.
One possible way to complete the canonical transformations is
\footnote{Different possibilities to complete the  canonical
transformations for remaining variables  will lead to  another
forms of dynamics, or
to equivalent form but in another frame of reference.}
\begin{equation}\label{Pip}
\begin{array} {ll}
\Pi_0=\frac{1}{2}(p_x^2-m^2)\,, \quad &X_0=\frac{x_0}{p_0}\,,\\
\Pi_i=p_i\,, \quad &X_i=x_i-\frac{p_i}{p_0}x_0\;.
\end{array}
\end{equation}
and the inverse transformation is
\begin{equation}  \label{piP}
\begin{array} {ll}
p_0=\sqrt{2\Pi_o+\vec\Pi^2+m^2}\,, \quad & x_0 = X_0\sqrt{2\Pi_o+\vec\Pi^2
+m^2}\,,\\
p_i=\Pi_i\,,    \quad                    & x_i=X_i+\Pi_iX_0\;.
\end{array}
\end{equation}
In terms of the new variables the total Hamiltonian is
\begin{equation} \label{HT}
H_T=e\Pi_o+\lambda p_e\;.
\end{equation}
and the equations of motion separate into two parts; one for
the canonical pairs    \( (e, p_e)\) and \(X_0, \Pi_0\),
with dependence from the Lagrange multiplier \(\lambda(\tau)\)
\begin{eqnarray} \label{XPep}
 \dot{X}_0=e\,,    &\quad & \dot{e}=\lambda \,,\\
\dot{ p}_e=-\Pi_0\,,  &\quad &  \dot{\Pi}_0 =0 \,,
\end{eqnarray}
constrained by
\(
 \Pi_0= 0
\)
and the equations of motion for the variables  $(X_i,\Pi_i)$
\begin{equation}      \label{PHUR}
\dot{X}_i=0\, \quad \quad   \dot{\Pi}_i=0\,,
\end{equation}
which have a unique solution with initial values free from
any restriction.
One can construct the reduced Poincare -Cartan 1-form for physical
unconstrained variables $X_i,\Pi_i$ from
(\ref{P-Cp}),   rewritten in terms of the new canonical
variables
\begin{equation}
\Theta=\Pi_0dX^0-\Pi_idX_i + p_ede-(e\Pi_o+\lambda p_e)dt+
d(X_0(\Pi_0+m^2))\,,
\end{equation}
by considering the projection onto the constraint shell
\begin{equation}
\Theta^{\ast}=\Theta\bigl{|}_{ \Pi_0=0,~p_e=0}=-\Pi_idX_i+d(X_o)m^2\;.
\end{equation}
Thus we have convinced ourselves that the  variables $\Pi_i, X_i$ --
are Jacobi's coordinates for the obtained unconstrained
theory with zero Hamiltonian.
Now we shall show how
to reconstruct the unconstrained Hamiltonian in terms of
initial variables using the generating function to new set
of canonical pairs (\ref{Pip}) and Hamilton-Jacobi
equation.
To find the unconstrained system whose Jacobi's coordinates are
$\Pi_i,X_i$ let us write down
the generating function $S(\Pi,x)$ of the canonical transformation
$( x,p) \to (X,\Pi)$ (\ref{Pip})
\begin{equation}
p=\frac{\partial S(\Pi,x)}{\partial x}\,,\;\;\;\;\;\;
X=\frac{\partial S(\Pi,x)}{\partial \Pi}\;.
\end{equation}
One can easily  verify from the condition
\begin{equation}
\Pi dX-pdx=d(X_o(\Pi_o+m^2)),
\end{equation}
that  the  function
\begin{equation} \label{genfun}
S(\Pi,x)=x_0\sqrt{2\Pi_0+\vec \Pi^2+m^2}-x_i\Pi_i\;
\end{equation}
generates the above  canonical transformations (\ref{Pip}).
Restriction the generating function
by the condition  $\Pi_o=0$ leads to the function
\begin{equation}
S^{\ast}(\Pi_i,x_i, x_0) =S(\Pi,x)\bigl{|}_{\Pi_o=0}=
x_0\sqrt{\vec \Pi^2+m^2}-x_i\Pi_i\;,
\end{equation}
which we shall now treat as generating function defined on
the unconstrained phase space \( (x_i, p_i)\) and depended explicitly
on some parameter  $x_0$, which has the  meaning of evolution parameter
for the obtained reduced system.
To verify this, one can use the generating function
$S^{\ast}(\Pi_i,x_i, x_0) $ to
write down the inverse transformation  for variables in
the  reduced Poincare -Cartan form  directly on the constraint shell
\begin{equation}
\Theta^{\ast}=-\Pi_idX_i+m^2dX_o|_{\left\{
\begin{array}{l}
X_i=\frac{\partial S^{\ast}}{\partial \Pi_i}=x_i\\
p_i=\frac{\partial S^{\ast}}{\partial x_i}=\Pi_i
\end{array}\right.}=-p_idx_i+\sqrt{\vec p^2+m^2}dx_o\;.
\end{equation}

From this form it follows that we get the Hamiltonian system
for a relativistic particle
\begin{eqnarray}
\frac{dx_i}{dt}&=&\{x_i,h\}=\frac{2p_i}{\sqrt{\vec p^2+m^2}} \\
\frac{dp_i}{dt }&=&\{p_i,h\}=0 \,,
\end{eqnarray}
in the instant form of the dynamics with the parameter $t:=x_0$
and the Hamiltonian  defined from the reduced generating function
\begin{equation}
h=: \frac{\partial S^\ast}{\partial x_0} = \sqrt{\vec{p}^2 + m^2}.
\end{equation}


\section{Hamiltonian reduction of FRW cosmological models}


\subsection{Scalar field with minimal coupling to gravity}

After performing the Legendre transformation
on the Lagrangian in the action (\ref{tru}) describing the
 dynamics of a homogeneous scalar field with
minimal coupling to FRW space time one finds that
the phase space spanned by the canonical pairs
\((a, p_a), (N_c,P_N)\) and
\((\Phi, P_\Phi )\) is restricted by the
primary  constraint
\begin{equation}
P_N = 0
\end{equation}
and secondary constraint
\begin{equation} \label{LPS1}
C=\frac{\kappa p_a^2}{12}+\frac{3ka^2}{\kappa r_0^2}
-\frac{P_\Phi^2}{2a^2}\;.
\end{equation}
Exploiting the nondegenerate character of the metric
\( (a\neq 0) \) the secondary constraint (\ref{LPS1}) can be rewritten
in the equivalent form
\begin{equation} \label{tilcon}
\tilde C=a^2C =a^2\left(\frac{\kappa p_a^2}{12}+\frac{3ka^2}{\kappa r_0^2}
\right)
 -P_\Phi^2/2\;,
\end{equation}
which shows the separability of the gravitational and the
matter source part in constraint.
To obtain the reduced Hamiltonian describing the
evolution of cosmic scalar factor \(a\)
one can introduce the new  canonical coordinates  for scalar
field
\begin{equation}
\Pi_\Phi:=P_\Phi^2/2\;,\;\;\;\;
T_\Phi :=\Phi/P_\Phi \,.
\end{equation}
After this redefinition the corresponding Poincare-Cartan form
\begin{equation} \label{pcm}
\Theta= p_a da + \Pi_\Phi dT_\Phi  -\frac{N}{a^2}\tilde C
dt +d\left(\Pi_\Phi T_\Phi \right),
\end{equation}
projected onto the constraint shell reduces to
\begin{equation} \label{rpcm}
\Theta^\ast = p_a da + H(a) dT_\Phi  +
d\left(H(a)T_\Phi \right),
\end{equation}
where the reduced Hamiltonian that governs the scale factor \(a\)
evolution in time \( T_\Phi\) is
\begin{equation}
H(a):=
a^2\left(\frac{\kappa p_a^2}{12}+\frac{3ka^2}{\kappa r_0^2}\right)\,.
\end{equation}

Note, that there is another possibility to reduce the theory.
The reduced theory can be formulated in terms of a scalar field.
To find the dynamics of the scalar field we perform the canonical
transformation on the scale factor
\begin{eqnarray} \label{mtime}
&&\Pi_a:=
a^2\left(\frac{\kappa p_a^2}{12}+\frac{3ka^2}{\kappa r_0^2}\right)\\
&&T_a:=
\int\limits_{a_o}^{a}{a^2da}\left(\frac{\kappa}{3}\Pi_a
-ka^4r_o^{-2}\right)^{-1/2}
\end{eqnarray}
and  as a result  the reduced  Poincare-Cartan form
in terms of scalar fiel variables
is
\begin{equation} \label{rpcmsc}
\Theta^\ast = P_\Phi d\Phi - H(P_\Phi) dT_a +
d\left(S(a, \Pi_a)- T_a\Pi_a \right),
\end{equation}
where the reduced Hamiltonian that describes the
evolution of
scalar field \(\Phi\) in time  \( T_a\) is
\begin{equation}
H(P_\Phi):=  \frac{1}{2}P_\Phi^2 \,,
\end{equation}
and the function \(S(a, \Pi_a)\) is the generating function of
the canonical transformation (\ref{mtime}).

\subsection{Scalar field with conformal  coupling to gravity}

In the case  of a homogeneous scalar field
conformally coupled to the  FRW space time (\ref{eq:lagrFWW+MS})
the phase space spanned by the canonical pairs
\((a, p_a), (N_c,P_N)\) and
\((\varphi, p_\varphi )\) is restricted by
the  primary   constraint
\begin{equation}
P_N = 0\,,
\end{equation}
and secondary constraint
\begin{equation}\label{scc}
C := \Pi_\varphi -\Pi_a\,,
\end{equation}
where
\begin{equation}
\Pi_a:= \frac{\kappa p_a^2}{12}+ \frac{3ka^2}{\kappa r_o^2} \,,
\quad \quad
\Pi_\varphi:= \frac{p_\varphi^2}{2}+\frac{k\varphi^2}{2r_o^2}\,.
\end{equation}
The total Hamiltonian \( H_T := N_c C + \lambda(t) P_N \)
contains the arbitrary function \(\lambda(t) \) and thus
the Hamilton-Dirac equations
\begin{equation}
\begin{array}{l}
\dot a=-N_c\kappa p_a/6\\
\dot p_a=N_c6ka/(\kappa r_o^2)\end{array}\;\;\;\;
\begin{array}{l}
\dot N_c=\lambda\\
\dot P_N = C\end{array}\;\;\;
\begin{array}{l}
\dot \varphi=N_cp_\varphi\\
\dot p_\varphi=-N_ck\varphi/r_o^2\;
\end{array}
\end{equation}
cannot be solved in a unique way.
According to the scheme described in the preceding sections
to implement the Hamiltonian reduction one
can search for a transformation to a new set
of canonical variables in terms of which
the equations of motion separate into independent parts:
the physical (independent of the
arbitrary function) and the unphysical one with unpredictable
evolution.
To achieve this let  us perform the canonical transformation
from \((p_a\,,a)\) and \((p_\varphi, \varphi)\)
to the new canonical pairs such that matter part of the constraint
\(\Pi_\varphi\)
becomes one of the new  canonical momenta
\begin{eqnarray} \label{P}
\Pi_\varphi=
\frac{p_\varphi^2}{2} +\frac{k\varphi^2}{2r_o^2}\,.
\end{eqnarray}
Using the generating function
\begin{equation}
S(\Pi_\varphi, \varphi):=
\int\limits_{a_o}^{a}{da}\sqrt{2\Pi_\varphi
-\frac{k}{2r_o^2}\varphi^2}\,,
\end{equation}
the corresponding  canonical conjugated coordinate \(T_\varphi\)
is
\begin{equation}
 T_\varphi=
\int\limits_{a_o}^{a}\frac{da}{\sqrt{2\Pi_a
-\frac{k}{2r_o^2}\varphi^2}}\,
\end{equation}
and the reduced action reads
\footnote{
Based on this action one can  derive the Hubble parameter
\(
H^2=
\frac{1}{a^4(T_c)}\frac{\kappa}{3} \Pi_\varphi
-\frac{k}{r_o^2a^2(T_c)}\;
\)
and convince ourselves that it  corresponds to
the radiation dominated Fridmann model
with the constant \( \Pi_\varphi \).}
\begin{eqnarray}
W^\ast [a] =
\int p_ada+(\frac{\kappa}{12}p_a^2+\frac{3k}{\kappa r_o^2}a^2)
dT_\varphi +
d\left(S(\Pi_\varphi, \varphi) -  \Pi_\varphi T_\varphi\right),
\end{eqnarray}

It is worth  mentioning that if instead of matter part the gravitational
part of constraint \(\Pi_a\)
will be used for the construction of the new
canonical momenta
then the reduced action describing the evolution of scalar
field is
\begin{eqnarray}
W^\ast[\varphi]&=&\int p_\varphi d\varphi
-\frac{1}{2}(p_\varphi^2+
\frac{k\varphi^2}{r_o^2})dT_\varphi\;.
\end{eqnarray}

\subsection{Spinor field as source field for FRW Universe}

The Hamiltonian reduction of this model is achieved
along  the same lines as
in the previous section. However, dealing with  fermion fields
there are  some specific features
due to the presence of the second class constraints.

The  action (\ref{S}) for the homogeneous spinor field in FRW Universe
is degenerate and the corresponding primary constraints are
\begin{equation}  \label{CONSpin}
\begin{array}{ccl}
C_N &:=&p_N=0\\
C_\psi&:=&p_\psi+\frac{i}{2}\bar\psi\gamma^o=0\\
C_{\bar\psi}&:=&p_{\bar\psi}+\frac{i}{2}\gamma^o\psi =0\,.
\end{array}
\end{equation}
They satisfy the algebra
\begin{equation}  \label{eq:alg1}
\{C_N,C_\psi\}=0 \;,\;\;\;\;\; \{C_N,C_{\bar\psi}\}=0\;,\;\;\;\;\;
\{C^{(1)}_\psi,C^{(1)}_{\bar\psi}\}=-i\gamma^o \,.
\end{equation}

According to the Dirac prescription the evolution in time
is governed by the total  Hamiltonian
\begin{equation}
H_T=H_c+\lambda_NC_N+C_\psi\lambda_\psi+
\lambda_{\bar\psi}C_{\bar\psi}\,,
\end{equation}
with the arbitrary functions \(\lambda \) and  the canonical
Hamiltonian \( H_c\)
\begin{equation} \label{HC}
H_{c}=N_c
\left[-\left(
\frac{\kappa p_a^2}{12}+
\frac{3ka^2}{\kappa r_o^2}
\right) +a{\cal H}_D\right]\,.
\end{equation}
The requirement to  conserve the  constraints during the evolution
fixes the  functions $\lambda_\psi$ and $\lambda_{\bar\psi}$
\begin{equation} \label{lspin1}
\lambda_{\bar\psi}=i N_cma\bar\psi\gamma^o, \;\;\;\;\;\;\;\;\;
\lambda_{\psi}=i N_cma\gamma^o\psi \,.
\end{equation}
but leaves  the function \(\lambda_N\) unspecified it
and leads to the existence of the secondary constraint
\begin{equation}
C := \frac{\kappa p_a^2}{12}+
\frac{3ka^2}{\kappa r_o^2}
-a{\cal H}_D =0 \,.
\end{equation}
Due to the algebra of constraints (\ref{eq:alg1}) and the Poisson brackets
of secondary constraint \( C \) with any other
\begin{equation} \label{PB}
\{C ,C_\psi\}=ma\bar\psi\;,\;\;\;\;
\{C, C_{\bar\psi}\}=-ma\psi \;,\;\;\;\;
\{C, C_{N}\}=0\;,
\end{equation}
one can verify that no additional constraints emerge.
\footnote{
Secondary constraint $C$ is conserved in weak sense
\begin{equation}
\dot C =i Nm p_a(C_\psi\gamma^o\psi+
\bar\psi\gamma^oC_{\bar\psi})\approx 0.
\end{equation}
}
The algebra  (\ref{eq:alg1}) and (\ref{PB}) shows that the constraints
represent a mixed system of first and second class constraints.
In order to perform the Hamiltonian reduction we will start with
rewriting the constraints into  an equivalent form
such that the first class constraints form the ideal of algebra
and the algebra of second class constraints is canonical.
This  equivalent set of constraints
$\bar C_\psi, \bar C_{\bar\psi}$
is given in the Appendix B.
The canonical character of the new algebra
$\{\bar C_\psi,\bar C_{\bar\psi}\}=-1$
allows to perform the canonical transformation  that converts
the new second class constraints $\bar C_\psi,\bar C_{\bar\psi}$
to the pair of canonical variables
\begin{equation} \label{NEWCON}
\begin{array}{l}
\bar \Pi_\psi=\bar
C_\psi\,,\\ \bar Q_\psi=\bar C_{\bar\psi}\,,
\end{array}\;\;\;\;
\begin{array}{l}
\Pi_\psi=i p_\psi\gamma^o+\frac{1}{2}\bar\psi \,,\\
Q_\psi=p_{\bar\psi}-\frac{i}{2}\gamma^o\psi \,.
\end{array}
\end{equation}
This means that the  dynamics of phase space
variables \(\bar Q_\psi, \bar \Pi_\psi\) is completely
``frozen'' and other canonical pairs  change in time independently of them.
In other words we can  everywhere in the formulas omit this variables
without destroying the dynamics of the physically relevant quantities.
Turning  to the reduction due to the first class constraints
let us pass to the new Hamiltonian constraint ${\cal{C}}$
\begin{equation}  \label{eq:newconstr}
{\cal{C}}: = \frac{1}{a} C =
\frac{1}{a}\left(\frac{\kappa p_a^2}{12}+
\frac{3ka^2}{\kappa r_o^2}\right)
-i m\Pi_\psi\gamma^oQ_\psi\,,
\end{equation}
assuming that the metric is nondegenerate $a\neq 0$.
In order to achieve the reduction
for first class constraint we perform  the canonical transformation
from the \((p_a, a)\) to the new variables \(( \Pi_a,Q_a) \)
such that
\begin{equation} \label{Pq}
\Pi_a= \frac{1}{a}\left(\frac{\kappa p_a^2}{12}+
\frac{3ka^2}{\kappa r_o^2}\right)\,.
\end{equation}
Using the generating function
$S(a,\Pi_a)$
\begin{equation}
S(a,\Pi_a)=\frac{6}{\kappa}\int\limits_{a_o}^{a}da
\sqrt{
\frac{\kappa}{3}a\Pi_a
-ka^2r_o^{-2}}\,,
\end{equation}
one can find  the variable
canonically  conjugated to  $\Pi_a$
\begin{equation} \label {timePar}
T_a=\int\limits_{a_o}^{a}{ada}\left(
\frac{\kappa}{3}a\Pi_a
-ka^2r_o^{-2}
\right)^{-1/2} \,,
\end{equation}
and after  projection onto the  constraint shell
\(
{\cal C}=0, \quad \bar C_\psi= 0, \quad  \bar\Pi_\psi=0,
\quad
\bar C_{\bar\psi}=0, \quad \bar Q_\psi=0
\)
the reduced action is
\begin{equation}
W^\ast[Q_\psi]= \int dQ_\psi\Pi_\psi + m \Pi_\psi\gamma^o Q_\psi dT_a\,.
\end{equation}
Thus we have derived the  standard  Dirac
Hamiltonian   for reduced spinor field
and this matter source corresponds to the
case of the dust filled Universe; the
Hubble constant  behaves as
\begin{equation}
H^2=\left(\frac{1}{a}\frac{da}{dT_a}\right)^2
=\frac{\kappa}{3}\frac{M_D}{a^3}-\frac{k}{a^2r_o^2}
\end{equation}
with the constant $M_D$.

We shall finish with one remark concerning the simple
generalization of the above result to a more complex system.
It is interesting to note that if one  includes the interaction of
massive spinor with the scalar massless one in the action  of the
following type
\begin{eqnarray} 
W[ g, \Phi,\Psi ] &=& \int\limits_{}^{}d^4x\sqrt{-g}\left[
-\frac{1}{16\pi G}\,
{}^{(4)}R + \frac{1}{12}\, {}^{(4)}R\Phi^2 + \frac{1}{2}
g^{\mu\nu}\partial_\mu\Phi\partial_\nu\Phi\right.\\
&&\left.+\frac{i}{2}\left(\bar\Psi\gamma^\mu(x)\bigtriangledown_\mu\Psi-
\bigtriangledown_\mu\bar\Psi\gamma^\mu(x)\Psi\right)
-m\bar\Psi\Psi-\mu\Phi\bar\Psi\Psi\right]\,,\nonumber
\end{eqnarray}
then the action obtained after supposition of the FRW Universe
\begin{eqnarray} 
W[a, N_c, \varphi,\psi]& = &\int dt\left[-
\frac{3}{\kappa}\frac{\dot a^2}{N_c}
+\frac{1}{2}\frac{\dot \varphi^2}{N_c}
+\frac{i}{2}(\bar \psi\gamma^o\dot\psi-\dot{\bar\psi}\gamma^o\psi)
\right.\\
&&\left.
-N_c\left(-\frac{3}{\kappa}\frac{ka^2}{r_o^2}
+\frac{k\varphi^2}{2r_o^2}
+(ma+\mu\varphi)\bar\psi\psi\right)
\right]\,,\nonumber
\end{eqnarray}
can be connected  with the  action describing the interaction of
fermion field and massless scalar field.
Let us consider two possible cases.

a). \underline{ $\kappa m^2 < 6\mu^2$ }.
One can convince ourself that after introduction
of the new scalar field $\phi$ and the scale factor $\alpha$
\begin{equation}
ma+\mu\varphi=\mu\phi\sqrt{1-\frac{\kappa}{6}\frac{m^2}{\mu^2}}
\;;\;\;\;a+\frac{m}{\mu}\frac{\kappa}{6}\varphi=
A\sqrt{1-\frac{\kappa}{6}\frac{m^2}{\mu^2}}
\end{equation}
we get the action  for the massless spinor interacting with
the field $\phi$
\begin{eqnarray}  \label{eq:ac1}
W[A, N_c, \phi,\psi]& = &\int dt\left[-
\frac{3}{\kappa}\frac{\dot A^2}{N_c}
+\frac{1}{2}\frac{\dot \phi^2}{N_c}
+\frac{i}{2}(\bar \psi\gamma^o\dot\psi-\dot{\bar\psi}\gamma^o\psi)
\right.\\
&&\left.
-N_c\left(-\frac{3}{\kappa}\frac{kA^2}{r_o^2}
+\frac{k\varphi^2}{2r_o^2}
+\tilde\mu\phi\bar\psi\psi\right)
\right],\nonumber
\end{eqnarray}
and the new coupling constant
\begin{equation}
\tilde\mu=\mu\sqrt{1-\frac{\kappa}{6}\frac{m^2}{\mu^2}}\;.
\end{equation}

b). \underline{  $\kappa m^2>6\mu^2$}.
In this case one can use another transformation,
\begin{equation}
\varphi=\frac{1}{\sqrt{1-\frac{6}{\kappa}\frac{\mu^2}{m^2}}}
\left(\phi-\frac{6}{\kappa}\frac{\mu}{m}A\right)\;;\;\;\;\;\;
a=\frac{1}{\sqrt{1-\frac{6}{\kappa}\frac{\mu^2}{m^2}}}
\left(A-\frac{\mu}{m}\phi\right)\,,
\end{equation}
and get the action
\begin{eqnarray} \label{eq:ac2}
W[A, N_c, \phi,\psi]& = &\int dt\left[-
\frac{3}{\kappa}\frac{\dot A^2}{N_c}
+\frac{1}{2}\frac{\dot \phi^2}{N_c}
+\frac{i}{2}(\bar \psi\gamma^o\dot\psi-\dot{\bar\psi}\gamma^o\psi)
\right.\\
&&\left.
-N_c\left(-\frac{3}{\kappa}\frac{kA^2}{r_o^2}
+\frac{k\phi^2}{2r_o^2}
+\tilde m A\bar\psi\psi\right)
\right],\nonumber
\end{eqnarray}
with the new mass for the fermion field
\(\tilde m=m\sqrt{1-\frac{6}{\kappa}\frac{\mu^2}{m^2}}\;.
\)
One can verify that these  two actions are related by
the field redefinition
\begin{equation}
A\;\to\;i\frac{\kappa}{6}\phi ,
\quad \quad
\phi\;\to\;i \frac{6}{\kappa}A,
\end{equation}
and thus it is enough to reduce one of the actions
(\ref{eq:ac1}),(\ref{eq:ac2}).

For the action (\ref{eq:ac1})
the energy constraint
\begin{equation}
C= -
\frac{\kappa p_a^2}{12}-
\frac{3ka^2}{\kappa r_o^2}
+\frac{p_\phi^2}{2} +\frac{k\phi^2}{2r_o^2} +
\tilde\mu \phi{\cal H}_D\,,
\end{equation}
again has separable contributions from the
gravitational and the  matter part.
After introduction of the new canonical momentum
\begin{equation}
\Pi_a:=
\frac{\kappa p_a^2}{12}-
\frac{3ka^2}{\kappa r_o^2}
\end{equation}
and the corresponding conjugated coordinate \(T_a\) in the same
manner as  for the case of the conformal scalar field
the following action for the physical scalar and spinor fields
can be derived
\begin{equation}
W^\ast[\phi, \psi]=
\int  dQ_\psi\Pi_\psi + p_\phi d\phi -H dT_a,
\end{equation}
with physical Hamiltonian describing the
system of interacting spinor and scalar fields
\begin{equation}
H:= \frac{p_\phi^2}{2} + \frac{k\phi^2}{2r_o^2}+
\tilde\mu \phi{\cal H}_D\,.
\end{equation}


\section{Classical and quantum observables for FRW Universe}



\subsection{Extended quantization: Wheeler-deWitt equation}

According to the Dirac prescription in the extended quantization scheme
one considers the classical constraints to be the conditions on
the state vector $\Psi$
\cite{Wheel}, \cite{deWitt} ,
\footnote{The standard procedure of letting
$P_N \rightarrow -i\partial_N,\,
p_a \rightarrow -i\partial_a, \,
P_\Phi \rightarrow -i\partial_{P_\Phi} $
is assumed.}
\begin{eqnarray} \label{eq:WdW}
&&P_N \Psi =0\,,\\
&&H_T\Psi =0 \,.
\end{eqnarray}
Quantum observable in this  quantization scheme
are constructed with analogy to that of the two-dimensional relativistic
spin zero bosonic  Klein-Gordon field  as expectation value
\begin{equation}
\bigl< O \bigl> = \int d\phi\left(
\Psi^\ast O \partial_a (\Psi) -  \partial_a (\Psi^\ast) O \Psi^\ast
\right)\,.
\end{equation}
However, as it has been analyzed by Kaup and Vitello
\cite{KaupVitello} this conventional interpretation
cannot be used without  violating the correspondence principle.
More precisely, it has  been shown that the expectation values for
the scalar fields and the cosmic scale factor do not correspond to the
classical values; their evolution describe  the expansion phase of
Friedmann evolution, but then instead  of contraction, the
expectation values tunnel through the barrier and continue to expand.
Below  it will be demonstrated  that opposite to
this situation the canonical quantization of the unconstrained system
obtained  in the preceding part
of the paper leads directly to the fulfillment of
the correspondence principle.

\subsection{Reduced quantization: Heisenberg equation}

To analyze the correspondence principle let us consider the case of a
conformal scalar field in the closed Friedmann Universe.
As it has been shown the
evolution of scale factor $a$ in conformal time $t$
is governed by the harmonic oscillator Hamiltonian which
after conventional quantization  reads
\begin{equation}
\hat H=\frac{\kappa}{12}\hat p^2+
\frac{3}{\kappa r^2_o}\hat a^2\;.
\end{equation}
Assuming the quantum state in the  form
\begin{equation}
\Psi=\frac{1}{(\alpha^2\pi)^{1/4}}
\exp\left[\frac{i}{\hbar}p_oa-\frac{(a-a_o)^2}{2\alpha^2}\right]
\end{equation}
where $a_o$ and $p_o$ are the mean values of the coordinate and the
momentum
respectively (real parameter $\alpha$ characterizes the mean square
deviation of $a$) and using the solution of Heisenberg equations
for the operators $\hat a(t)$ and $\hat p(t)$
\begin{eqnarray}
&&\hat a(t)=\hat a(0)\cos\frac{t}{r_o}
-\frac{\kappa r_o}{6}\hat p(0)\sin\frac{t}{r_o}\,,\\
&&\hat p(t)=\frac{6}{\kappa r_o}\hat a(0)\sin\frac{t}{r_o}
+\hat p(0)\cos\frac{t}{r_o}\;,
\end{eqnarray}
one can  find the time dependence of the mean values of $\hat a(t)$ and
$\hat p(t)$,
\begin{eqnarray}
&&\overline{a(t)}=\int\limits_{-\infty}^{+\infty}
\Psi^*\hat a(t)\Psi da=a_o\cos\frac{t}{r_o}
-\frac{\kappa r_o}{6}p_o\sin\frac{t}{r_o}\,,\\
&&\overline{p(t)}=\int\limits_{-\infty}^{+\infty}
\Psi^*\hat p(t)\Psi da=\frac{6}{\kappa r_o}a_o\sin\frac{t}{r_o}
+p_o\cos\frac{t}{r_o}\;.
\end{eqnarray}
This means that we have the correspondence with the
classical formulae
\begin{eqnarray}
&&a(t)=r_o\sqrt{\frac{\kappa}{3}|H|}\sin\left(\frac{Q-T_c}{r_o}\right)\\
&&p(t)=\sqrt{\frac{12}{\kappa}|H|}\cos\left(\frac{Q-T_c}{r_o}\right)
\end{eqnarray}
when constants are taken  as
\begin{equation}
\overline{a(0)}=a_o=
r_o\sqrt{\frac{\kappa}{3}|H|}\sin\frac{Q}{r_o}\;,\;\;\;
\overline{p(0)}=p_o=\sqrt{\frac{12}{\kappa}|H|}\cos\frac{Q}{r_o}\;.
\end{equation}

At the end we note that there is no wave packet diffusion when
the mean  square deviation
\begin{equation}
\overline{(\triangle a(t))^2}=\frac{\alpha^2}{2}
\left(\cos^2\frac{t}{r_o}+\left(\frac{\kappa r_o}{6}\right)^2
\frac{\hbar^2}{\alpha^4}\sin^2\frac{t}{r_o}\right)
\end{equation}
is time independent. This  holds for the special value of $\alpha$
\begin{equation}
\alpha^2=\hbar\frac{\kappa r_o}{6}\;.
\end{equation}

\section{Concluding remarks}

In the present paper the method of Hamiltonian reduction for
reparametrization invariant mechanical systems have been elaborated.
This approach is based on the choice of adapted coordinates
using the generating function of the canonical transformation
that is a solution of the corresponding  Hamilton -- Jacobi equation.
We have derived the  reduced Hamiltonians for the Friedmann
cosmological models
with  homogeneous scalar and spinor field matter sources and
find the corresponding observable time.
The obtained reduced Hamiltonians have two attractive peculiarities:

i. They are  the generators
of evolution with respect to observable time;

ii. They are conserved quantities which can be treated as
the energy of the reduced systems.\footnote{
The conservation of the conformal matter Hamiltonian with respect
to conformal time translations follows from conformal symmetry of the
Robertson -- Walker space-time.}

Furthermore, the representation for the Hubble parameter and the red shift is founded
in terms of the Dirac observables in the frame of the generalized
Hamiltonian dynamics and correspondence between field Friedmann models
and perfect fluid Friedmann models with different equations of state
has been established.

The comparison of extended quantization with the Wheeler-deWitt equation
and the canonical quantization of unconstrained system
shows the conceptual advantage of later.
In reduced system with the Schr\"odinger type equation instead of the
to the Wheeler-deWitt equation the wave function is normalizable
and has clear standard quantum mechanical interpretation.
It is shown that quantum observables treated as expectation values of
the Dirac observables properly describe the original classical theory.

It is in order here to make a remark concerning the relation to the conventional
gauge-fixing method. Certainly, the results derived in the present note
by reduction without introduction of  gauge
functions can be reproduced by the gauge fixing
method.
However, from our derivation it is clear that due to the complicated
relations between the initial variables  and the observable time
the gauge functions depend on the initial
variables in a complex way which is  difficult to guess.

Finally we would like to point out the possibility to exploit the
suggested approach.
The method elaborated in the present article can be used
in the description of the other cosmological models, like Bianchi cosmologies,
with different type of global symmetries.
But the applicability  of obtained results to a general problem
of observables meets with the several difficulties. Nevertheless,  we hope that
in combination with other refined methods our approach will help to extend our
understanding of the puzzle of the observables in theory of gravity.

\section{ Acknowledgments}

We would like to thank S.Gogilidze, D.Mladenov, H.-P.Pavel  and
V.Pervushin for
helpful and critical discussions.
This work was supported in part by the Russian Foundation of Basic
Research, Grant No 99\--01\--00101.

\appendix

\section{Dirac equation in FRW space time}
\label{appA}
To describe a spinor field on a Rimanian manifold the
vierbein fields $h^\mu_a(x) \,\, \mu, \nu, a, b = 0,1,2,3 $
$$
ds^2=g_{\mu\nu}dx^\mu dx^\nu=\eta_{ab}(h^a_\mu dx^\mu)(h^b_\nu dx^\nu)\;;\;
\quad \eta_{ab}:=(+---),
$$
and the Dirac $\gamma$-matrices with a specific dependence on space time
coordinates are introduced
$$
\gamma^\mu(x)=h^\mu_a(x)\gamma^a .
$$
The following relations between the vierbein fields and
the metric tensor  $g_{\mu\nu}$ hold
\begin{equation}
h^\mu_ah_{b\mu}=\eta_{ab}\;;\;\;h^a_\mu h_{a\nu}=g_{\mu\nu}\;;\;\;
h^a_\mu h^\mu_b=\delta^a_b\;;\;\;\;h^a_\mu h^\nu_a=\delta^\nu_\mu\;.
\;;\;\;\;
h_{a\mu}=\eta_{ab}h^b_\mu=g_{\mu\nu}h^\nu_a\;.
\end{equation}
The Dirac equation for spinors in  curved space time reads
\begin{equation} \label{DEQ}
(i\gamma^\mu(x)\bigtriangledown_\mu-m)\Psi(x)=0\;,
\end{equation}
with the  covariant derivative
\begin{equation} \label{COVDER}
\bigtriangledown_\mu\Psi(x)=[\partial_\mu+\frac{1}{4}C_{abc}h^c_\mu
\gamma^b\gamma^a]
\Psi(x)\;,
\end{equation}
where Ricci coefficients
\begin{equation}
C_{abc}\equiv (\bigtriangledown_\mu h^\nu_a)h_{b\nu}h_c^\mu\;;\;\;\;\;
\bigtriangledown_\mu h^\nu_a=(\Gamma^\nu_{\mu\lambda}
-h^\nu_b\partial_\mu h^b_\lambda)
h^\lambda_a\;,
\end{equation}
are introduced. For the specific case of the Robertson -- Walker metric,
\begin{equation}  \label{TFRN}
ds^2=a^2(t)\tilde ds^2=
a^2(t)\left[(N(t)dt)^2-\left(1+\frac{kr^2}{4r_o^2}\right)^{-2}
\left(dr^2+r^2(d\xi^2+\sin^2\xi d\zeta^2)\right)\right]\,,
\end{equation}
the following  vierbein fields
\begin{equation} \label{tetr}
\left\{
\begin{array} {ll}
h^{\underline o}_o=aN&h^{\underline 1}_1=a\left(1+\frac{kr^2}{4r_o^2}
\right)^{-1}\\
h^{\underline 2}_2=ar\left(1+\frac{kr^2}{4r_o^2}\right)^{-1}&
h^{\underline 3}_3=ar\sin \zeta\left(1+\frac{kr^2}{4r_o^2}\right)^{-1}
\end{array}\right.
\end{equation}
are used in the main text. Here
the vierbein indices are underlined. The Dirac equation then
looks
\begin{eqnarray}
\frac{i}{a}\left[\gamma^o\frac{1}{N}\frac{\partial}{\partial t}
+\gamma^1\left(1+\frac{kr^2}{4r_o^2}\right)\frac{\partial}{\partial r}
+\gamma^2\frac{1+\frac{kr^2}{4r_o^2}}{r}\frac{\partial}{\partial \zeta}
+\gamma^3\frac{1+\frac{kr^2}{4r_o^2}}{r\sin \zeta}
\frac{\partial}{\partial \xi}\right.\nonumber\\
\left.+\frac{3\dot a}{2aN}\gamma^o+\frac{1-\frac{kr^2}{4r_o^2}}{r}\gamma^1
+\frac{\cot \zeta}{2r}\left(1+\frac{kr^2}{4r_o^2}\right)\gamma^2
\right]\Psi(x)-m\Psi(x)&=&0\;.
\end{eqnarray}
To maintain the space homogeniety of the Friedmann Universe
we suppose that the spinor field is only time dependent.
In the main text the
FRW Universe with the spinor matter source is formulated in terms
of the fermion  variable $\psi$
\begin{equation}
\psi(t)=a^{3/2}(t)\Psi(t).
\end{equation}


\section{ Separation of first and second class constraints
in model with spinor field}


The set of constraints $C_A=(C_{\psi}, C_{\bar\psi}, C )$ represent
a mixed system of first  and second class constraints; the rank of the
Poisson matrix \( {\cal M}= \{C_A,C_B\} \) is equal to two.
The explicit form of the Poisson  matrix is
$$
{\cal M} =
\left(\begin{array}{cc}
\triangle&K\\
-K^T&0
\end{array}\right)\,,
$$
where and $\triangle $ and $K$ denote
$$
\triangle=\left(\begin{array}{cc}
0&-i\gamma^o\\
-i\gamma^o&0
\end{array}\right)\;\;\;\;\;\;\;\;
K=\left(\begin{array}{c}
-ma\bar\psi\\
 ma\psi
\end{array}\right)\,.
$$
In order to perform the reduction procedure it is useful to
separate first and second class constraints. One can easily  verify that
applying the similarity  transformation \(T\)
$$
T=
\left(\begin{array}{cc}
1&0\\
K^T\triangle^{-1}&1
\end{array}\right)\,,\;\;\;\;\;\;\mbox{Sdet}T\neq 0
$$
to the  constraints $C_A$
$$
\tilde C =T\cdot C =
\left(\begin{array}{ccc}
1&0&0\\
0&1&0\\
i ma\gamma^o\psi&-i am\bar\psi\gamma^o&1
\end{array}\right)\cdot
\left(\begin{array}{c}
C_\psi\\
C_{\bar\psi}\\
C\end{array}\right)
$$
we achieve the separation of the constraints  on the surface
defined by the second class constraints
$$
\{\tilde C ,\tilde C_\psi\}=i ma
\tilde C_\psi\gamma^o\;\;\;\;\;
\{\tilde C,\tilde C_{\bar\psi}\}=-i ma
\gamma^o\tilde C_{\bar\psi}\,.
$$
To have this separation on the whole phase space one can pass
to the new set of constraints
\begin{eqnarray} \label{eqconsf}
\bar C &=&\tilde C + ma\tilde C\psi \tilde C_{\bar\psi}\nonumber\\
&=&\Pi+ ma \left(p_\psi
p_{\bar\psi}-\frac{i}{2}
[p_\psi\gamma^o\psi+\bar\psi\gamma^op_{\bar\psi}]\right)
-\frac{1}{4}a{\cal H}_D,\\
\label{gr}
\bar C_\psi&=&-i \tilde C_\psi\gamma^o=-i p_\psi\gamma^o+\frac{1}{2}
\bar\psi \,, \\
\bar C_{\bar\psi}&=&\tilde C_{\bar\psi}=
p_{\bar\psi}+\frac{i}{2}\gamma^o\psi \,.
\label{sconstr}
\end{eqnarray}
In this new set $\bar C$ belongs to the ideal of the algebra
of constraints
\begin{equation}
\{\bar C,\bar C_\psi\}=\{\bar C_N,\bar C_{\bar\psi}\}=0\,,
\end{equation}
and second class constraints $\bar C_\psi,\bar C_{\bar\psi}$
obey  the canonical algebra
\begin{equation}
\{\bar C_\psi,\bar C_{\bar\psi}\}=-1\,.
\end{equation}

\section{
Reduced Hamiltonian as conserved quantity from conformal symmetry}


In this Appendix we discuss the existence of
time independent reduced Hamiltonians from the  geometrical
standpoint.
The  Friedmann -- Robertson -- Walker space-time is conformally
flat
\begin{equation}
ds^2_{FRW}=A^2(x)ds^2_{Minkowski}\,.
\end{equation}
In the flat Friedmann Universe the conformal factor $A(x)$ is simple
scale factor $a(T_c) $
and it is easy to verify that the conformal time translation
is a conformal symmetry
\begin{equation}
\pounds_{\partial_{T_c}}g_{\mu\nu}=
\pounds_{\partial_{T_c}}(a^2(T_c)\eta_{\mu\nu})=
\eta_{\mu\nu}\partial_{T_c}a^2(T_c)=
g_{\mu\nu}2\frac{\dot a}{a}.
\end{equation}
It is well-known that if  space time possesses
the  conformal Killing vector and
matter energy-momentum tensor is traceless, then one can construct the
conserved quantity as follows.
Considering  the covariant derivative of contraction of the
stress tensor and the confomal Killing vector
\begin{eqnarray}
\bigtriangledown_\mu P^{\mu}&=&\bigtriangledown_\mu\left(\xi_\nu
T^{\mu\nu}\right)=
\xi_\nu\bigtriangledown_\mu T^{\mu\nu}+T^{\mu\nu}\frac{1}{2}
(\bigtriangledown_\mu\xi_\nu+\bigtriangledown_\nu\xi_\mu)\nonumber\\
&=&\xi_\nu\bigtriangledown_\mu T^{\mu\nu}+\frac{1}{n}T^\mu_\mu
\bigtriangledown^\nu\xi_\nu \nonumber\,,
\end{eqnarray}
and assuming the covariant conservation of
the traceless matter energy-momentum tensor
\begin{equation}
\bigtriangledown_\mu T^{\mu \nu}\ = 0\,, \quad \quad T^\mu_\mu=0\,,
\end{equation}
we have the conservation law
for four-vector \(P^{\mu}\) in the
covariant differential form
\begin{equation}
\bigtriangledown_\mu P^{\mu}=0.
\end{equation}
To get the global conserved quantity one can
integrate this equality over the whole space-time and
use Gauss theorem
\begin{eqnarray} \label{charge}
&&\int\limits_{V}^{}d^4x\sqrt{-g}\bigtriangledown_\mu\left(\xi_\nu
T^{\mu\nu}\right)=
\int\limits_{V}^{}d^4x\frac{\partial}{\partial x^\mu}
\left(\sqrt{-g}\xi_\nu T^{\mu\nu}\right)\\
&&=\int\limits_{T_{c}'}^{}(\xi_{T_c})_\nu T^{\mu\nu}\sqrt{-g}d^3x-
\int\limits_{T_{c}''}^{}(\xi_{T_c})_\nu T^{\mu\nu}\sqrt{-g}d^3x\nonumber\,,
\end{eqnarray}
where in the last line we specify the Killing vector corresponding
to the  conformal translation in Robertson -- Walker space-time.
For the conformal scalar field with Lagrangian
\begin{equation}
{\cal L}=\sqrt{-g}\left(\frac{1}{2}g^{\mu\nu}\partial_\mu\Phi
\partial_\nu\Phi
+\frac{1}{12}\ {}{(\!4\!)}\!\!R\Phi^2\right)\,,
\end{equation}
the canonical stress tensor
\begin{equation}
T^{C}_{\mu\nu}=\partial_\mu\Phi\partial_\nu\Phi
-g_{\mu\nu}\frac{1}{\sqrt{-g}}{\cal L}
\end{equation}
has nonzero trace $T^{{C} \mu}_{\mu}\neq 0$.
However, according to \cite{ChernikovTag},
 one can pass to the improved tensor
\begin{equation}
T_{\mu\nu}=T^{C}_{\mu\nu}-
\frac{1}{6}\left[-\ {}{(\!4\!)}\!\!R_{\mu\nu}+
\partial_\mu\partial_\nu-g_{\mu\nu}\partial^\mu\partial_\mu\right]\Phi^2\,,
\end{equation}
which is traceless \( T^\mu_{\mu}=0.\)
Thus for a conformal time Killing vector in adapted coordinates
$\xi_{T_c}=(1,0,0,0)$
and for homogeneous scalar field
$
\varphi(T_c)=a(T_c)\Phi(T_c)
$
from eq. (\ref{charge}) it follows that
\begin{equation}
H=\int\limits_{T_{c}}^{}(\xi_{T_c})_o T^{oo}\sqrt{-g}d^3x=
V_{(3)}\left(\frac{p_\varphi^2}{2}+\frac{k\varphi^2}{2r_o^2}\right)
\end{equation}
is conserved charge that coincides with the
reduced Hamiltonian derived in the main
text.


\begin{thebibliography}{99}
\bibitem{Bergmann} P.Bergmann, Rev.Mod.Phys. {\bf 33}, 510 (1961)
%
\bibitem{KucharTime} K.Kuchar,
The Problem of Time in Canonical Quantization of Relativistic
Systems in  {\it Conceptual Problems of Quantum
Gravity \/} ed. by A.Ashtekar, J.Stashel.
(Birkhauser, Boston, 1991).
%
\bibitem{HajicekTime} P.Hajicek,  Nucl. Phys. B (Proc.Suppl.)
{\bf 57}, 115 (1997).
%
\bibitem{Dir58} P.A.M.~Dirac. Proc.Roy.Soc., {\bf A~246} (1958) 333;
Phys.Rev. {\bf 114}, 924 (1959).
%
\bibitem{ADM} R.Arnowitt, S.Deser and  C.W.Misner. in {\it Gravitation :
An Introduction to Current Research}, ed. L.Witten,
(Wiley: New Yourk 1962), 227.
%
\bibitem{Ryan}{M. Ryan, {\it Hamiltonian Cosmology},
Lecture Notes in Physics N 13,
(Springer-Verlag, Berlin, 1972).}
%
\bibitem{ReggeTeit}T. Regge and C. Teitelboim, Ann. Phys.
{\bf 88 }, 286 (1974).
%
\bibitem{Faddeev} L.D.Faddeev, Usp. Fiz. Nauk {\bf 136},  437 (1982).
%
\bibitem{Torre} G.Torre, Phys.Rev., {\bf 48}, R2373 (1993).
%
\bibitem{DiracL} P.A.M. Dirac,
{\it Lectures on Quantum Mechanics.\/} Belfer Graduate School of Science,
(Yeshiva University, New York, 1964).
%
\bibitem{Sunder}
K. Sundermeyer, {\it Constrained Dynamics\/},
Lecture Notes in Physics N 169,
(Springer Verlag, Berlin - Heidelberg - New York, 1982).
%
\bibitem{HenTeit}
M. Henneaux  and C. Teitelboim,
{\it Quantization of Gauge Systems \/},
(Princeton University Press, Princeton, NJ, 1992).
%
%
\bibitem{KucharEmbed} K.Kuchar, Foundation of Physics,
{\bf 16}, 193 (1986).
%
\bibitem{HajicekTop} P.Hajicek, Phys. Rev. D {\bf 34} (1986) 1040;
J.Math.Phys.{\bf 30}, 2488  (1989).

\bibitem{Wolf} J. Wolf, {\it Spaces of Constant Curvature\/},
           (University of California, Berkley, 1972).
%
\bibitem{Bernui} A.Bernui, G.I.Gomero, M.J.Reboucas, and A.F.F.Teixera,
         CBPF-NF-051/98, gr-qc/9903037.
%
\bibitem{GKP} S.A. Gogilidze, A.M.Khvedelidze, and V.N. Pervushin
J.Math.Phys. {\bf 37}, 1760 (1996);
Phys.Rev. D {\bf 53}, 2160 (1996).
%
\bibitem{MTW} C.W. Misner, K.S. Thorne, and J.A. Wheeler,
{\it Gravitation} (Freeman, San Francisco, 1973).
%
\bibitem{Deser} S. Deser,  Phys.Let {\bf 134}, 419 (1984).
%
\bibitem{Beken}
J.D. Bekenstein, Ann.of Physics {\bf 82}, 535 (1974).
%
\bibitem{CalColJack} C.G. Calan, S. Coleman, and R.Jackiw,
Ann.Phys {\bf 59}, 42 (1970).
%
\bibitem{Page}D.N. Page, J.Math.Phys. {\bf 32}, 3427 (1991).
%
\bibitem{StarkovichCooperstock}
S.P. Starkovich, and F.I. Cooperstock,  Astrophys. J. {\bf 398}, 1 (1992).
%
\bibitem{Isham} C.J. Isham, and J.E. Nelson, Phys.Rev. D {\bf 10}, 3226
        (1974).
%
\bibitem{Zanelli} T. Christodoulakis, and J. Zanelli,
                  Phys. Rev. D {\bf 29}, 2738 (1984).
%
\bibitem{Levi-Civita} T. Levi-Civita, and U. Amaldi,
{ \it Lezioni di Meccanica razionale\/}, (Nicola Zanichelli, Bologna, 1927).
%
\bibitem{Shanmugadhasan} S. Shanmugadhasan, J. Math. Phys {\bf  14\/}, 677
(1973).
%
\bibitem{Wheel} J.A.Wheeler. In Batelle Recontres :
1967 Lectures  in Mathematics and Physics, edited by  C. DeWitt and
J.A.Wheeler, Benjamin, New York, (1968).
%
\bibitem{deWitt} B.S.DeWitt. Phys.Rev. {\bf 160},  1113 (1967).
%
\bibitem{KaupVitello}D.J.Kaup and A.P.Vitello, Phys. Rev.
D {\bf 9},  1648 (1974).
%
\bibitem{ChernikovTag} N.A. Chernikov and E.A. Tagirov,
         Ann. Inst. H. Poincare, {\bf 9A} 109 (1968).
\end{thebibliography}
\end{document}